\documentclass[a4paper,
               biblatex,     
               keeplastbox,   
               ]{jacow}
\usepackage{pdfpages,multirow,ragged2e} %
\makeatletter%
	\ifboolexpr{bool{xetex}}
	 {\renewcommand{\Gin@extensions}{.pdf,%
	                    .png,.jpg,.bmp,.pict,.tif,.psd,.mac,.sga,.tga,.gif,%
	                    .eps,.ps,%
	                    }}{}
\makeatother

\ifboolexpr{bool{xetex} or bool{luatex}} 
 {}                                      
 {\usepackage[utf8]{inputenc}}           

\usepackage[USenglish]{babel}

\usepackage{amssymb}

\ifboolexpr{bool{jacowbiblatex}}%
 {%
  \addbibresource{THPR36.bib}
 }{}
\newcommand{\fig}[1]{Fig.~\ref{#1}}
\newcommand{\eqn}[1]{Eq.~\ref{#1}}
\newcommand{\code}[1]{\texttt{#1}}

\begin{document}

\title{Automated Anomaly Detection on European XFEL Klystrons}

\author{A. Sulc, Helmholtz-Zentrum Berlin für Materialien und Energie, Berlin, Germany \\
A. Eichler, T. Wilksen, Deutsches Elektronen-Synchrotron, Hamburg, Germany}
	
\maketitle

\begin{abstract}
High-power multi-beam klystrons represent a key component to amplify RF to generate the accelerating field of the superconducting radio frequency  (SRF) cavities at European XFEL. Exchanging these high-power components takes time and effort, thus it is necessary to minimize maintenance and downtime and at the same time maximize the device's operation. In an attempt to explore the behavior of klystrons using machine learning, we completed a series of experiments on our klystrons to determine various operational modes and conduct feature extraction and dimensionality reduction to extract the most valuable information about a normal operation. To analyze recorded data we used state-of-the-art data-driven learning techniques and recognized the most promising components that might help us better understand klystron operational states and identify early on possible faults or anomalies.
\end{abstract}

\section{Introduction}
European XFEL is currently operating 25 klystrons. They play a crucial role in the acceleration and operation of the European XFEL. 
They function as radio frequency (RF) power amplifiers, providing high-power RF signals that are used to accelerate the charged particles to high energies as they travel through the linear accelerator structure. 
The performance and reliability of the klystrons directly impact the accelerator's ability to deliver the desired beam energy, intensity, and stability to the experiment. 
Klystrons that are not functioning optimally can lead to fluctuations in the RF power, resulting in beam energy variations, reduced beam intensity, or even beam loss. 
Failure of a klystron can cause a complete interruption in the accelerator's operation, as the loss of RF power will prevent the particle beam from being accelerated, leading to downtime for the accelerator facility and disrupting ongoing experiments. 

At the same time, klystrons are highly complex devices, with many interdependent components and parameters. Diagnosing and predicting potential failures in them can be very challenging because they have relatively high stability and mostly operate as black boxes from a signal viewpoint. 

In this work, we leverage machine learning techniques for anomaly detection to better understand klystron operational states and identify potential faults before they lead to downtime.
We then employ a specialized one-class anomaly loss~\cite{pmlr-v80-ruff18a} which is an unsupervised machine learning approach that reduces the dimensionality of the inputs to isolate key characteristics distinguishing normal functions. By training models on this reduced feature set, we can recognize anomalies and incipient failures based solely on klystron signals that we retrieve from DOOCS~\cite{rehlich2005status} and collect with DxMAF~\cite{schuette2023mo2bco02}, providing a data-driven means of maximizing uptime for these delicate and expensive devices.

\section{Related Work}
The first fully digital version of the Klystron Lifetime Management 
(KLM) system was developed by Butkowski et al.~\cite{butkowski2009klm} to maximize the lifetime of klystron tubes used in the X-Ray Free Electron Laser (XFEL) at DESY and has been successfully running since then. 
In their setting at European XFEL, klystrons are linear-beam vacuum tubes that operate at 1.3 GHz and 10 MW power to accelerate electron bunches for the European XFEL. 
A crucial component is the continuous operation of klystrons for at least 20 years, thus the klystron lifetime needs to exceed 60,000 hours. 
The KLM itself is a digital system that detects exceptional events like arcing, RF breakdowns, etc., and takes preventive actions (interrupt the drive) to avoid potential complications leading to longer downtime. 
In the implementation of~\cite{butkowski2009klm}, it is crucial to have a fast and reliable FPGA implementation that handles protection functions such as reflection limitation, forward/input power correspondence, and energy monitoring within 300 nanoseconds.
The KLM system serves as an effective preventive measure to avoid klystron damage from exceptional events during operation. Our research aims to go a step further;
by detecting potential issues at an even earlier stage before they manifest as urgent events, we can provide more lead time to anticipate problems and take preventive action.

In~\cite{sulc2023data} authors propose an anomaly detection approach using a neural network model to predict breakdowns on superconducting radio frequency (SRF) cavities at the European XFEL. They experiment with two models: one trained with a semi-supervised anomaly loss (SAL)~\cite{ruff2019deep} and another with binary cross-entropy loss (BCE). 
The SAL model uses a small set of labeled anomalous data along with a larger set of normal data, while the BCE model is trained in a fully supervised manner on the labeled data and requires balance in training data. 

Our approach is similar to~\cite{sulc2023data}, but we use the unsupervised one-class loss~\cite{pmlr-v80-ruff18a} (OC) instead of the SAL~\cite{ruff2019deep}, as we lack labeled anomalous cases. The model is trained solely on standard waveform modes, allowing it to learn normal patterns and detect deviations as anomalies without labels.
The one-class loss's sensitivity can potentially enhance the klystron lifetime and reliability by alarming operators based on training to sensitive modes of klystron signals characterizing the state. Additionally, we obtain a latent variable encoding the klystron state and potential drift(s) as the model's output embedding of the klystron state changes. 

\section{Method}
\subsection{One Class Loss}
Deep OC classification~\cite{pmlr-v80-ruff18a} is a loss function that aims to train a model to inputs belonging to a specific class(es) by learning solely from training data containing examples of that class.
This approach differs from traditional classification problems where the training data includes examples from all classes, and where the goal is to distinguish between the classes. The objective of OC loss is to model the characteristics of the single class of interest which is dominating the inputs, rather than distinguishing it from other classes.

Consider a function $f_\theta : \mathbb{R}^N\mapsto\mathbb{R}^M$ (a neural network). The objective of OC is to minimize the distance between a fixed (randomly chosen) hypersphere center $\mathbf{c}\in \mathbb{R}^M$ and the projection $f_\theta$ of most (non-anomalous points) $\mathbf{x}$,
\begin{equation}
    {\arg \min}_\theta \underset{\text{anomaly score }s\left(\mathbf{x}\right)}{\underbrace{\|f_{\theta}\left(\mathbf{x}\right)-\mathbf{c}\|_{2}}},
    \label{eq:oc}
\end{equation}

where parameters $\theta$ of $f$ are optimized to project $\mathbf{x}$ to $\mathbf{c}$. 
The OC loss is an anomaly score $s$. 
If $f_\theta$ has bias parameters, fix biases and $\mathbf{c}$ to avoid trivial solution~\cite{pmlr-v80-ruff18a}, where biases converge to $\mathbf{c}$.

Furthermore, the $f_\theta\left(\mathbf{x}\right)$ encodes $\mathbf{x}$'s state based on distance and position, thus outputs can be visualized and analyzed via T-SNE~\cite{van2008visualizing} as a state embeddings.

\subsection{Architecture}
The architecture consists of an input layer that takes a sequence of 205 features, followed by an LSTM layer with either 32 or 64 hidden units ($M$) to capture temporal dependencies in the input sequence. 
The LSTM layer's output is then mapped by a linear layer to real numbers, which is the final model embedding, see \eqn{eq:oc}.

\subsection{Inputs}
\label{sect:inputs}
 States of the klystrons are expressed by the following waveforms: \code{FD.F1} is forward power at the first klystron arm, \code{FD.F2} is forward power at the second klystron arm, \code{FD.FI} is forward power at klystron input, \code{FD.R1} is reflected power at first klystron arm, \code{FD.R2} is reflected power at second klystron arm, and \code{FD.RI} is reflected power at klystron input. Each waveform consists of 2048 amplitude and phase values. 

\subsubsection{Pre-processing}
\label{sect:inputs:pre}
First, we transform the amplitude and phase into in-phase and quadrature components (IQ).
Then we sub-sample waveforms, taking only every 10th value to lower computational requirements.
Our experiments show only \code{FD.RI} is essential without losing too much recognition ability.
Each waveform is normalized to zero mean and unit standard deviation.

\section{Results}
We selected two events that were detected post mortem as anomalous with our detection system, from which one led to a fatal failure of klystron (Feb. 1st 2024). 

\begin{figure}
    \centering
    \resizebox{1.0\linewidth}{!}{
    \begin{tabular}{c}
    \includegraphics[width=1.0\linewidth]{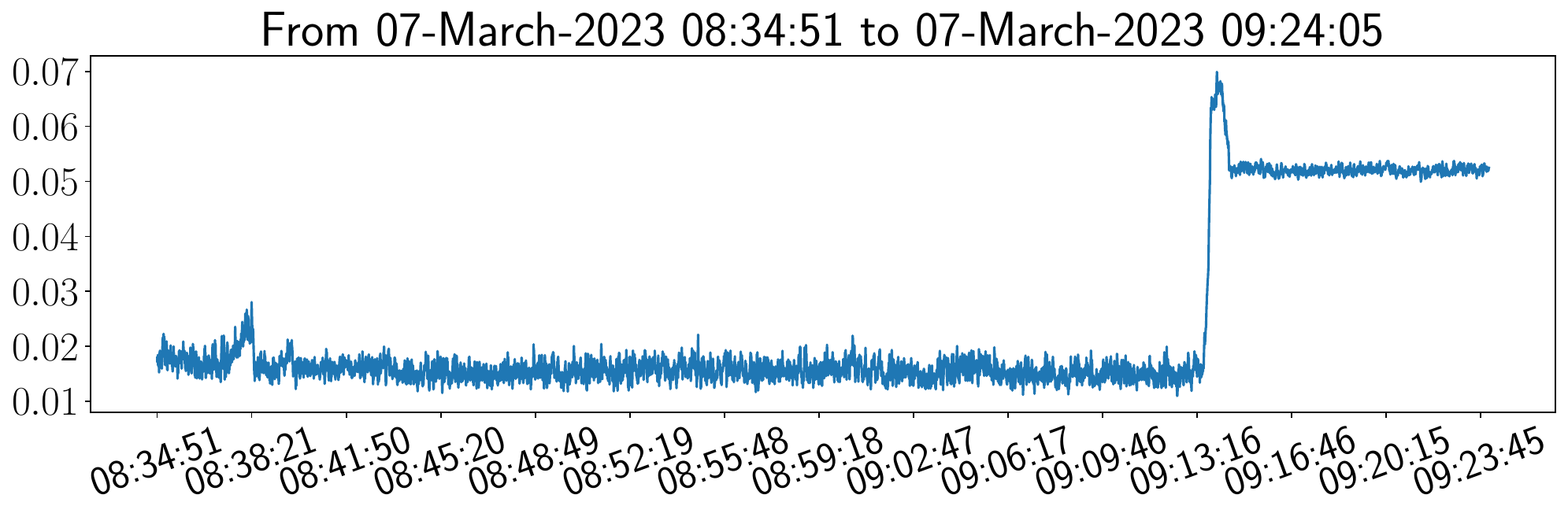}\\
    \includegraphics[width=1.0\linewidth]{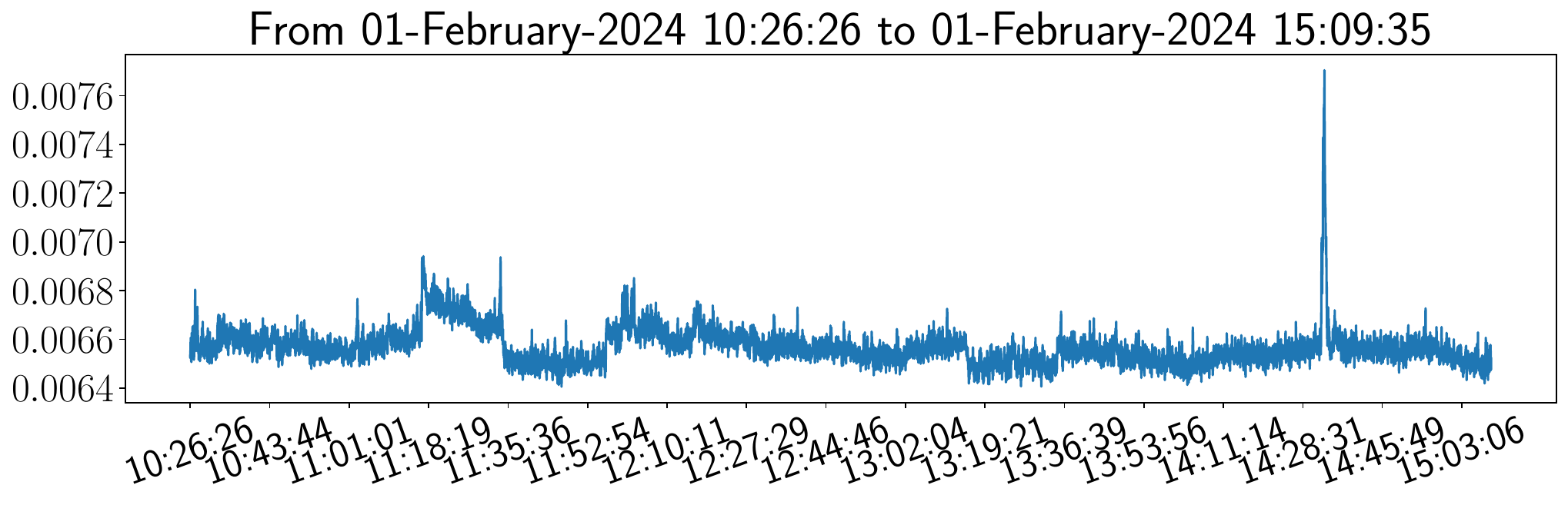}
    \end{tabular}}
    \caption{Anomaly scores $s$ of the event that took place on Mar. 7th 2023 (top) and Feb. 1st 2024 (bottom) 
    Top: Notice the bump around 9:13:16 which increases the $s$ until the klystron is shut down (after 9:23:45).
    Right: The anomaly score fluctuates quite significantly over an extended period (10:24:26 - 15:03:06). There is a peak around 14:30, which precedes severe disruption of observed signals.}
    \label{fig:s}
\end{figure}

\subsection{Event on Mar. 7th 2023}
The probable cause of this issue is multipacting, where electrons emitted from the cathode or other surfaces within the vacuum envelope of the klystron A13 can become trapped in a resonant trajectory, repeatedly striking the cavity walls.
This process can begin suddenly and escalate rapidly, creating a high density electron cloud that can absorb energy from the RF field, leading to excessive heating, outgassing, and ultimately RF breakdown in the klystron.

\begin{figure*}[t]
\begin{center}
    \includegraphics[width=1.0\linewidth]{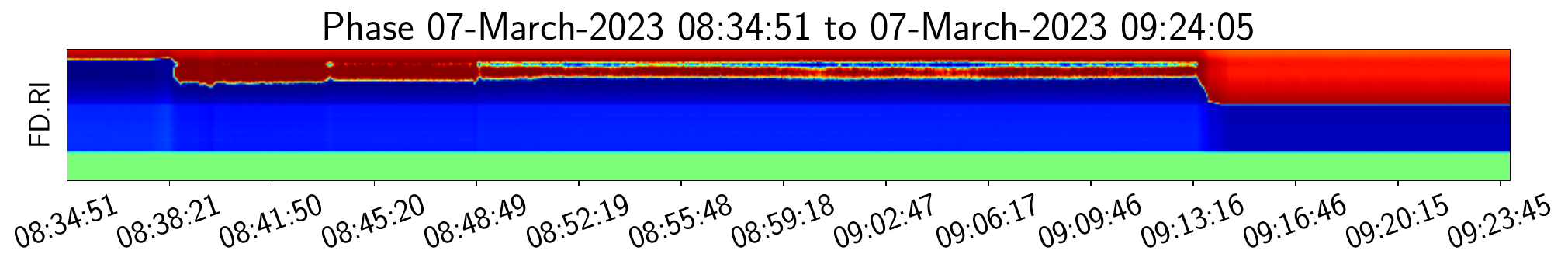}
\end{center}
\caption{The phase of the event that took place on Mar. 7th 2023. Each column is one waveform \code{FD.RI} of one pulse.}
\label{fig:AP20230307}
\end{figure*}

\begin{figure}[b]
    \centering
    \resizebox{1.0\linewidth}{!}{
    \begin{tabular}{cc}
    Mar. 7th 2023 & Feb. 1st 2024\\
    \includegraphics[width=0.5\linewidth]{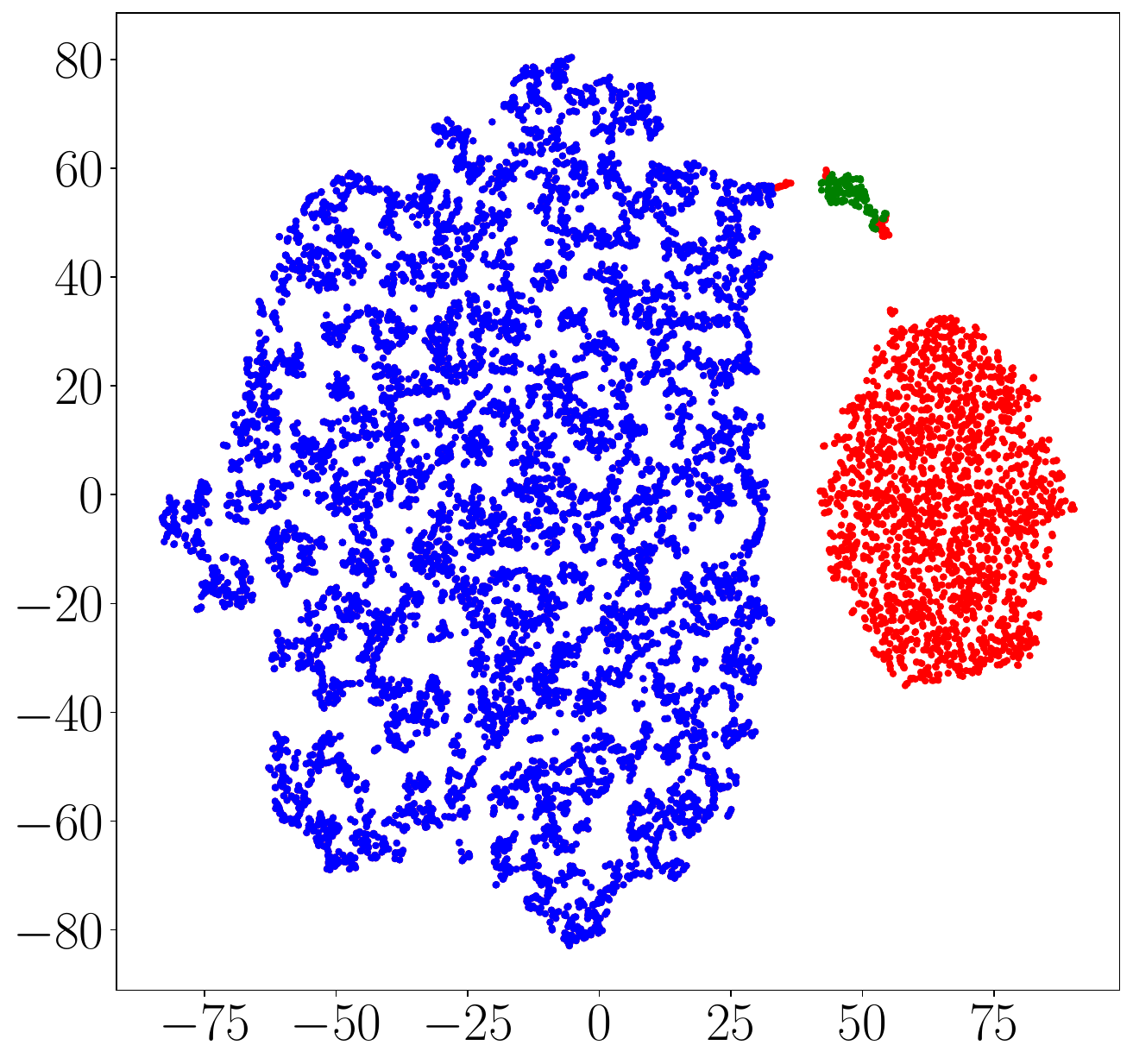} & \includegraphics[width=0.5\linewidth]{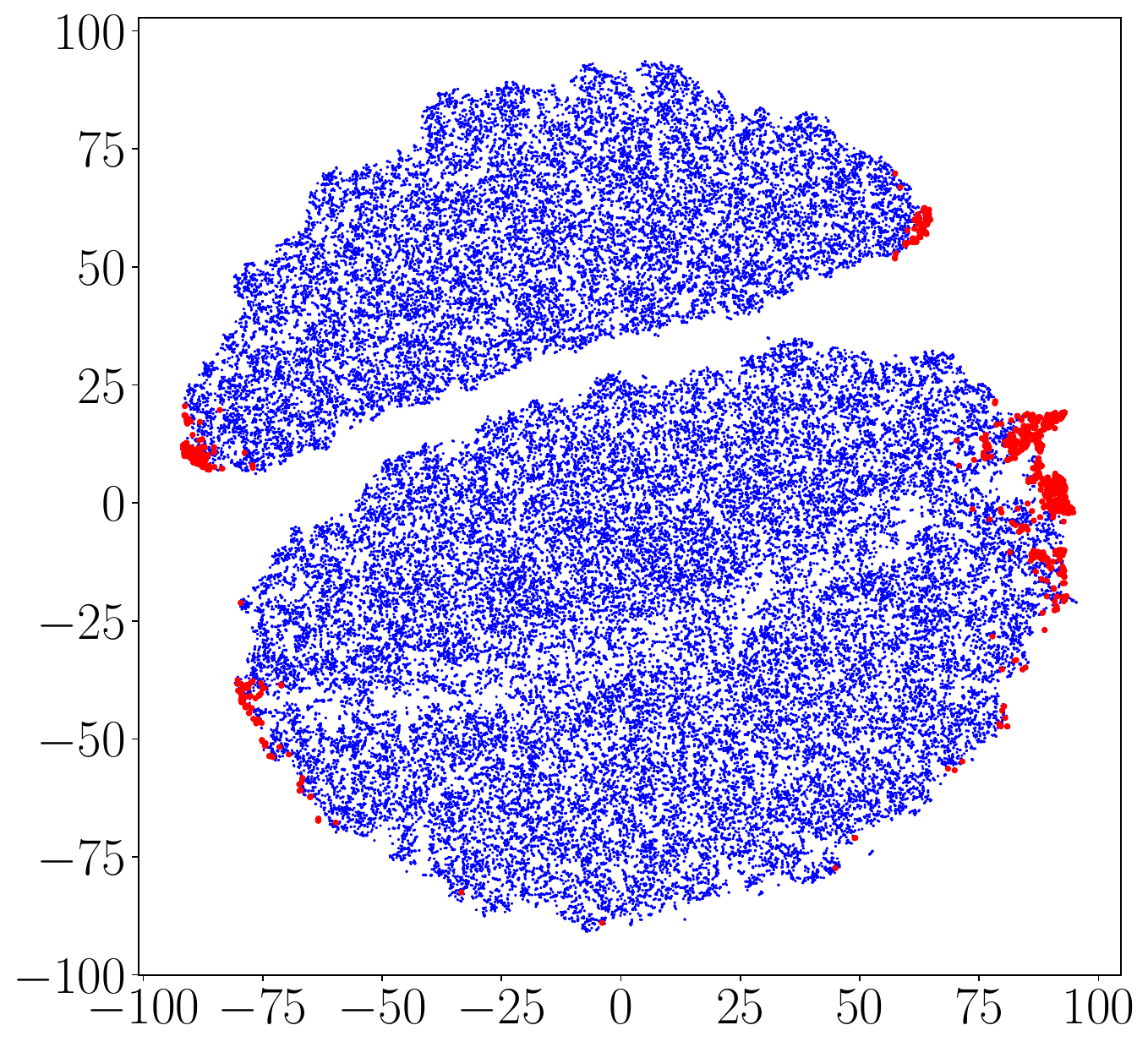}
    \end{tabular}}
    \caption{T-SNE Embedding shows a reduced projection of the network from $M$ dimensions onto 2D while distance is preserved. The colors of the left image encode different $s$ levels (red between $(0.04,0.06)$, blue between $(0,0.04)$ and green $(0.06,\infty)$. Similarly on the right figure, red color encodes $s$ above $0.007$.}
    \label{fig:TSNE}
\end{figure}

Our algorithm notices the first issue at 08:38:21, where there is a slight bump in $s$, see \fig{fig:s} (top). 
Furthermore, at around 9:13:16, there is a noticeable jump of $s$ followed by a steady increase in score.
An unexpected state change is observed in the \code{FD.FI} signal, as it is also visible from the phase in~\fig{fig:AP20230307} at that time. 
After this bump, which peaks approximately around 9:14:00, there is a noticeable change in the phase.
When we take a detailed look at the T-SNE embedding of the vectors $f$ in that time range, we can divide the space into three separate clusters according to \fig{fig:TSNE} (left): blue states corresponding to normal state(s), green points which are waveforms that are projected by $f$ away from the blue points, and red states which are states that follow after the peak (a klystron drift).

\begin{figure*}[t]
\begin{center}
    \includegraphics[width=1.0\linewidth]{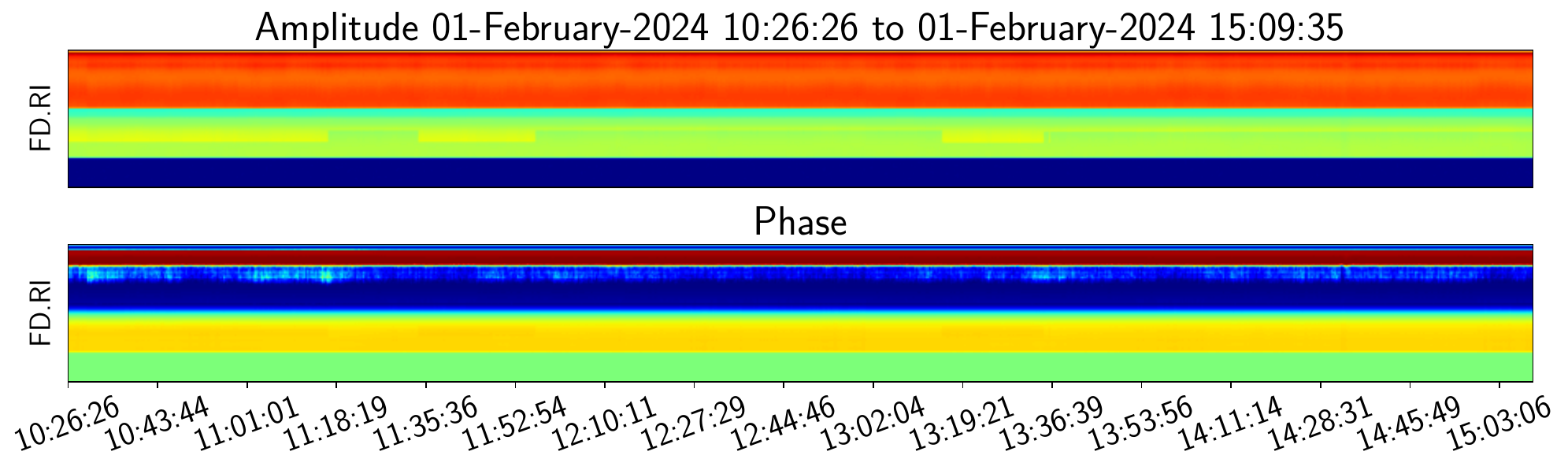}
\end{center}
\caption{Amplitude and phase of \code{FD.RI} of the event from Feb. 1st 2024.}
\label{fig:AP20240201}
\end{figure*}

\subsection{Event on Feb. 1st 2024}
The A16 klystron experienced a catastrophic failure, being unable to sustain the required high voltage, likely due to excessive dark current. The reflected power levels fluctuate erratically, another potential symptom of dark current issues. Despite attempts to mitigate arcing events, the arcing persisted, eventually leading to permanent damage to the klystron tube. Physical disassembly and inspection of internal components is required for comprehensive failure analysis after klystron replacement.

With the trained algorithm, one can observe mild fluctuations of $s$, see right \fig{fig:s}. For instance, approximately periods 11:18 - 11:35, 13:02 - 13:19, and finally a noticeable peak around 14:30 which might indicate a bigger problem before the station stops sending signals (after 15:09).
By looking at the input signals in~\fig{fig:AP20240201}, we can see that there are several mode changes (approximately 10:26 - 11:18, 11:35 - 11:52, 13:19 - 13:36), this is also visible in two distinct clusters in the T-SNE embedding in right \fig{fig:TSNE}. What is however quite noticeable is variation in the first third of the phase, where values fluctuate quite significantly.

\section{Conclusion}
In this work, we demonstrated the application of unsupervised deep one-class classification with an LSTM model for sequential anomaly detection in the operational signals of high-power klystrons at the European XFEL facility. 
By training solely on normal waveform data, the model learns to characterize standard klystron behavior and identify deviations as potential anomalies or faults.
We presented two case studies of actual events - a suspected multipacting issue on Mar. 7th 2023, and a catastrophic klystron failure on Feb. 1st 2024. 
Our algorithm identified precursors, flagging anomalous waveform patterns before the issues escalated to system failures or downtime which were assessed by the experts.
%

\section{Acknowledgements}
We acknowledge DESY (Hamburg, Germany) and HZB (Berlin, Germany), a member of the Helmholtz Association HGF, for their support in providing resources and infrastructure. 

We would like to thank all colleagues of the MCS and MSK groups, especially Julien Branlard, and importantly Vladimir Vogel and Chris Christou from the MHFp group for their contributions to this work, their help, and expertise.
%

%
%
\clearpage
\ifboolexpr{bool{jacowbiblatex}}%
	{\printbibliography}%
	{%
	
	
}
%
%
\end{document}